\documentclass[journal]{vgtc}                     

\onlineid{1122}

\vgtccategory{Research}

\vgtcpapertype{Theory/model}
\usepackage{caption,subcaption,enumitem}

\title{Heuristics for Supporting Cooperative Dashboard Design}
\author{%
  \authororcid{Vidya Setlur}{0000-0003-3722-406X},
  \authororcid{Michael Correll}{0000-0001-7902-3907},
  \authororcid{Arvind Satyanarayan}{0000-0001-5564-635X},
  and 
   \authororcid{Melanie Tory}{0000-0002-6806-9253}
}

\authorfooter{
  \item
  	Vidya Setlur is with Tableau Research.
  	E-mail: vsetlur@tableau.com
  \item
  	Michael Correll completed this work while at Tableau Research.
  	E-mail: m.correll@northeastern.edu
  \item Arvind Satyanarayan is with MIT CSAIL.
  	E-mail: arvindsatya@mit.edu
   \item Melanie Tory is with Northeastern University.
        E-mail: m.tory@northeastern.edu
}

\abstract{
Dashboards are no longer mere static displays of metrics; through functionality such as interaction and storytelling, they have evolved to support analytic and communicative goals like monitoring and reporting. Existing dashboard design guidelines, however, are often unable to account for this expanded scope as they largely focus on best practices for visual design. In contrast, we frame dashboard design as facilitating an \textit{analytical conversation}: a cooperative, interactive experience where a user may interact with, reason about, or freely query the underlying data. 
By drawing on established principles of conversational flow and communication, we define the concept of a \textit{cooperative dashboard} as one that enables a fruitful and productive analytical conversation, and derive a set of $39$ dashboard design heuristics to support effective analytical conversations. 
To assess the utility of this framing, we asked $52$ computer science and engineering graduate students to apply our heuristics to critique and design dashboards as part of an ungraded, opt-in homework assignment.
Feedback from participants demonstrates that our heuristics surface new reasons dashboards may fail, 
and encourage a more fluid, supportive, and responsive style of dashboard design. 
Our approach suggests several compelling directions for future work, including dashboard authoring tools that better anticipate conversational turn-taking, repair, and refinement and extending cooperative principles to other analytical workflows.
}


\teaser{
 \centering
  \includegraphics[width=\textwidth]{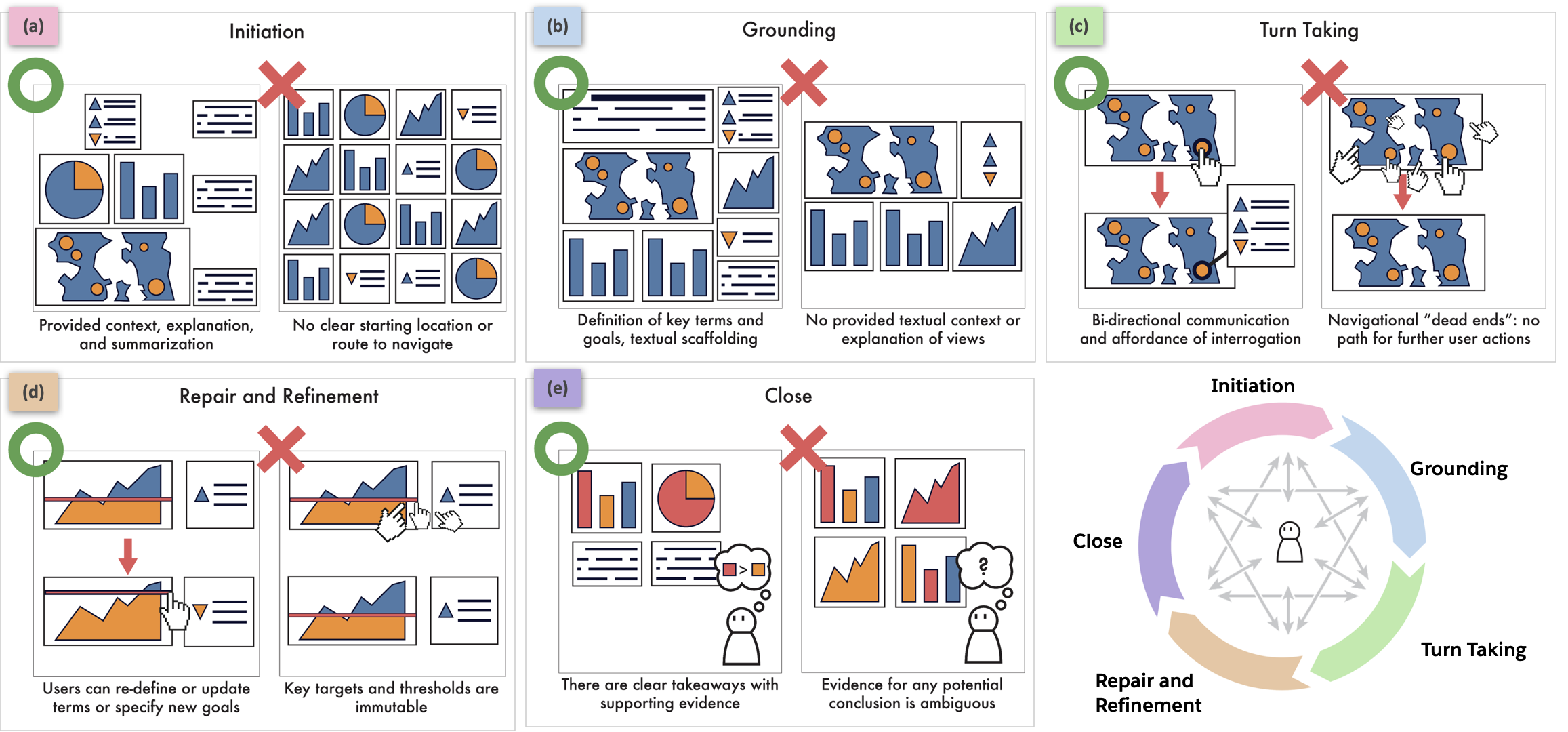}
  
  \caption{Successes and failures of \textit{cooperative dashboard} design throughout the five \change{analytic states} of a conversation \change{(a-e)}. Cooperative dashboards guide users through their data and, in contrast to static dashboards, provide bi-directional communication through interactivity to allow the user to change or refine their analytical goals, switch between topics of interest and levels of detail, correct or update the system if it provides irrelevant or incorrect information, and provide useful summaries of analytical actions. \change{Note that these conversation states are not necessarily sequential and the analyst can move between these various states.}}
\label{fig:teaser}
}

\graphicspath{{figs/}{figures/}{pictures/}{images/}{./}} 

\usepackage[dvipsnames,svgnames]{xcolor}

\usepackage{tabu}                      
\usepackage{booktabs}                  
\usepackage{lipsum}                    
\usepackage{mwe}                       
\usepackage{dialogue}
\usepackage{mathptmx}                  
\usepackage{subcaption}

\newcommand{\pheading}[1]{\vspace{4px}\noindent\textbf{#1}}

\definecolor{initiationcolor}{HTML}{f6b8d3}
\definecolor{groundingcolor}{HTML}{bdd7ef}
\definecolor{turntakingcolor}{HTML}{baeaaa}
\definecolor{repaircolor}{HTML}{eac6a1}
\definecolor{closecolor}{HTML}{b3a2de}

\newcommand{\change}[1]{\textcolor{black}{#1}}

\newenvironment{tight_itemize}{\begin{itemize} \itemsep -1pt}{\end{itemize}}

\begin{document}


\firstsection{Introduction}

\maketitle

\label{sec:introduction}
Dashboards have become ubiquitous for analyzing and communicating data because their expressive designs allow them to address a diverse range of purposes and contexts~\cite{sarikaya2018we}.
However, existing guidelines for dashboard design\,---\,whether in research or popular press~\cite{few:dashboarddesign, wexler2017big}\,---\,largely focus on issues of visual representation, perception, and graphic design. 
While important, this focus ignores the central role that interactivity and storytelling increasingly play in enabling users to explore, analyze, monitor, and track various data metrics~\cite{sarikaya2018we}.
What design guidance can we provide for these interactive capabilities?

\change{Prior work has described interaction as \emph{``engaging the data in dialogue''}~\cite{tominski2015interaction, dimara2019interaction, tory2005evaluating}\,---\,an analogy to human-human conversation that we find productive for thinking about what it means for a dashboard's interaction to be designed effectively. 
Just as a human conversationalist can be circumlocutory, obscurant, or rude, so too can interactions in a dashboard be repetitive, unclear, or user-unfriendly.
Moreover, gleaning insights from data is most productive and enjoyable when users can focus on answering the questions they have about their data rather than the mechanics of doing so~\cite{tory2019mean}.
But, what makes a dashboard an effective conversational partner?}

\change{We operationalize the conversation analogy by studying the \textit{pragmatics} of language use, or how language shapes meaning~\cite{saygin:2002}.
We define a dashboard as \textit{cooperative} if it facilitates an interactive loop that follows the \textit{Gricean Maxims}~\cite{Grice1975-GRILA-2}\,---\, influential work in pragmatics that assesses the quality of a cooperative, communicative interaction based on the quantity, quality, relation, and manner of information communicated. 
Moreover, we draw on work by Beebe et al.~\cite{beebe2004interpersonal} to model a cooperative analytical conversation as one where participants (in our case, the dashboard and the analyst) move between states of initiation, grounding, turn-taking, repair \& refinement, and close.}

Guided by these two frameworks, we enumerate a set of heuristics focused on \change{interactive, cooperative communication between a dashboard and a user.}
Through an iterative process with $16$ visualization practitioners, we distill down to a set of $39$ design heuristics for promoting the design of cooperative 
dashboard conversations. 
To evaluate the utility of these heuristics in practice, we conduct two exercises with $52$ computer science and engineering graduate students as part of optional, ungraded homework assignments. 
First, students were asked to use the heuristics to reflect on the efficacy of existing dashboard designs. 
Next, the students were asked to create a new dashboard or update the design of an existing dashboard based on the heuristics to better support cooperative conversational behavior with their target users. 

\change{Results of the classroom exercises indicate that our heuristics afford a new perspective for thinking about dashboard design.
While dashboards tend to be effective at initiation and grounding a conversation, they are often weaker with respect to turn-taking, repair \& refinement, and close. 
For instance, students noted how interactive results updating in place without any accompanying cues or messaging hinders turn-taking, as it can be difficult for a user to assess when an interaction is complete so they can resume their dialogue. 
When applying these heuristics to improve existing dashboard designs, students relied on textual annotation to provide contextual information and deliberately traded off visual aesthetics for clearer communication. 
Our results suggest opportunities for future work to study the impact of cooperative vs. uncooperative dashboard designs and to extend principles of cooperative conversation to analytical workflows beyond the dashboard.} 

\section{Related Work}
\label{sec:relatedwork}

Our work builds on three lines of research: understanding dashboard design and usage as representational media, conversational interactions with data, and design heuristics in HCI and visualization. 

\subsection{Understanding Dashboard Design and Usage}
Dashboards are pervasive. They operate as the primary portal to data for many people in work and daily life. Yet until recently, dashboards were given little attention by the visualization research community. A survey of dashboards in the wild~\cite{sarikaya2018we} offered a classification of dashboards and highlighted their criticality as a means of circulating data within organizations. An extension by Bach et al.~\cite{bach2022dashboard} identified six distinct dashboard genres and characterized content and composition design patterns.
Dimara et al. discussed the role dashboards play in supporting data-driven decision making~\cite{dimara2021unmet}, Zhang et al. described the work practices and challenges of dashboard creators~\cite{zhang2022visualization}, Lee-Robins and Adar~\cite{lee2022affective} characterized affective intents in visualizations and dashboards, and Tory et al.~\cite{tory2021finding} discussed the work practices of dashboard users. 
Research into dashboard design and construction includes approaches to enable layout and view consistency~\cite{qu2017keeping} and semantic snapping~\cite{kristiansen2021semantic}. Research into multiple coordinated views and composite visualizations is also relevant to dashboard design (for a survey see Roberts~\cite{roberts2007state} or Deng et al.~\cite{deng2022revisiting}). Dashboard design is typically a manual process that can be aided by design heuristics such as those introduced in this work. However, design heuristics may also be codified into systems that automatically generate dashboards (e.g.~\cite{key2012vizdeck}) or provide mixed-initiative support for dashboard creation~\cite{chen2020composition, wu2021multivision, pandey2022medley}. Our research focuses on better understanding the dialogue around dashboards by introducing a set of heuristics to support dashboard design and evaluation for analytical conversation.

\subsection{Conversational Interaction with Data}
The novelty in our heuristics stems from framing people's interaction with dashboards as a conversation. Designers have long recognized the power of interacting with computers in ways that emulate our conversational interactions with people. 
A long history of research on chatbots and other conversational interfaces is summarized in several surveys~\cite{agarwal2022chatbots, luo2022critical, chaves2021should, adamopoulou2020chatbots}. In recent years, this research theme has extended into interactions with data. A survey of natural language interfaces (NLIs) for data visualization was introduced by Shen et al.~\cite{Shen2022TowardsNL}. This body of work has led to an understanding of principles for cooperative communication design in conversational bots~\cite{chaves2021should, setlur:analyticalchatbots}, including behaviors such as communicability, conscientiousness, conciseness, manner, proactivity, and turn-taking (drawing on the Gricean Maxims~\cite{Grice1975-GRILA-2}).

Recent work recognizes that these cooperative principles apply beyond the scope of interfaces that employ spoken or written language. Most relevant are papers that characterize interactions with data and/or dashboards as data \emph{conversations}~\cite{fiore2015communication, muller2019data, tory2021finding}. Muller et al.~\cite{muller2019data} described how data science workers engage in back-and-forth interactions with data, especially for data wrangling. Tory et al.~\cite{tory2021finding} 
described how dashboards serve as a portal to data and a jumping-off point to further data activities. Their observation that dashboards alone are often ineffective in supporting these conversations, resulting in data being exported for use in spreadsheets, presentation tools, and reports, suggests a strong need for dashboards to evolve in ways that support more conversational forms of interaction. \emph{BOLT} explores the use of NLIs for dashboard authoring, wherein NL utterances are mapped to prevalent dashboard objectives to generate dashboard recommendations~\cite{bolt}.
Our work contributes a set of heuristics that can support designers in creating such cooperative, conversational dashboards and the systems that generate them.

\subsection{Heuristics in HCI and Visualization}
Our dashboard heuristics build upon a long history of design heuristics for interfaces and visualizations. In interface design and evaluation, perhaps the most well-known are Nielsen's~\cite{nielsen1994enhancing, nielsen2005ten} ten usability guidelines and Shneiderman et al.'s~\cite{shneiderman2016designing} eight golden rules. More specific heuristics have been developed for topics such as human-AI interaction~\cite{amershi2019guidelines}, augmented reality~\cite{endsley2017augmented, furmanski2002augmented}, and mobile computing~\cite{bertini2006appropriating}, among many others. Researchers have proposed numerous heuristics specific to conversational interaction with chatbots and voice assistants~\cite{langevin2021heuristic, sugisaki2020usability, hohn2020heuristic, maguire2019development, wei2018evaluating, fulfagar2021development, nowacki2020improving}.

Tory \& M\"{o}ller~\cite{tory2005evaluating} explored usability heuristics as a way to evaluate visualizations. 
Subsequently, there have been numerous efforts to develop and evaluate visualization-specific heuristics~\cite{forsell2010heuristic, vaataja2016information, tarrell2014toward, zuk2006heuristics, dowding2018development, cuttone2014four}.
The numerous high-level books, guidelines, and principles around dashboard design are also relevant (e.g.,~\cite{few:dashboarddesign, yigitbasioglu2012review, wexler2017big}), as are frameworks of user goals or intents that may help to guide visualization design (e.g., ~\cite{lam2017bridging,lee2022affective}), and design tools considered to support cognition~\cite{warebook:2008}. More recently, Lin et al.~\cite{lin:dashboardmining} introduced a data-driven approach for identifying a set of dashboard design rules from dashboards mined from the web. The rules describe view-wise relationships in terms of data, encoding, layout, and interactions and subsequently develop a recommender for dashboard design.

However, heuristics and guidelines for dashboards tend to focus on layout, structure, data and its visual representation, and usability. Our work augments these guidelines based on principles of cooperative conversation. The conversational framing offers a different perspective that aligns with an evolution of dashboards away from autocratic information artifacts and towards cooperative conversational partners.

\section{Analytic Conversation States}
\label{sec:analyticalconversation}
The motivation for this work stems from exploring how cooperative conversation guidelines for human-computer interfaces could inform the design and evaluation of interactive dashboards. Conversation is highly structured and organized according to set principles. Sacks et al.~\cite{sacks:1974} initiated the modern literature on conversational behavior by outlining a system of social interactions with specific properties. This interaction is characterized by a mechanism of exchange based on alternating dialogues of information. 

Beebe et al.~\cite{beebe2004interpersonal} break conversation down into five states (i.e., initiation, grounding, turn-taking, repair \& refinement, and close) that we adapt here for our discussion around interactive dashboards. {While Gricean Maxims~\cite{Grice1975-GRILA-2} provide guidelines for assessing the overall quality of a conversation, the conversation states specifically help define how an analytical conversation progresses through different interaction states; they also help organize the heuristics. We maintain, as per Tory et al.~\cite{tory2021finding}, that the users of dashboards are similarly engaged in ``data conversations'', so conversational structures (and pitfalls) can apply to dashboards and to considerations for their design. In this section, we introduce and apply these conversational states to dashboard interaction for supporting analytical conversation with the user (\autoref{fig:teaser}).



\subsection{Initiation}

Initiation is the first stage of conversation and requires one to be open to interacting with the other conversational participant(s). 
Greetings such as, ``Hello!" and ``How nice to see you!" are common ways to set the tone to welcome further dialogue. Conversations can also be initiated without any preliminaries using utterances such as, ``when will it stop raining?'' or including vocative or attention-seeking utterances such as, ``excuse me'' or ``hey!''

With respect to dashboard design, initiation can be thought of as both the state of the dashboard when the user first interacts with it, as well as any tutorials, explanations, or other tools for orienting the user to the dashboard's contents. Dhanoa et al.~\cite{dhanoa2022onboarding} suggest an ``onboarding model'' for new users of dashboards. A successful onboarding process, per this model, is mindful of the target user, the dashboard components that will need likely explanation, how these explanations will be serviced, and how this onboarding process connects to later patterns of usage. The means and goals of onboarding are then connected with an ``onboarding narrative''. For instance, a ``depth-first narrative'' might involve a serial explanation of every dashboard component (and their subcomponents) in detail. As in \autoref{fig:teaser}, a successful initiation in dashboard design provides the user with \change{\textbf{information and explanations} of components, but also clear options} for where to begin to understand their data. A failure can occur either through the lack of appropriate onboarding (e.g., an insufficient quantity of onboarding for the user, insufficient relevance to their task, or missing context) or even by presenting a ``data deluge'' of too many unconnected or unstructured views without a clear reading order or spatial organization.

Other strategies for successful initiation are to provide users with \change{\textbf{curated information and metadata}}. For instance, as in Srinivasan et al.~\cite{srinivasan2018augmenting}, dashboards can be augmented with ``data facts'' of potentially important relationships or patterns in the data. Or, as Gebru et al.~\cite{gebru2021datasheets}, a ``datasheet'' or important context and metadata could be provided to a user prior to any analysis.

\subsection{Grounding}
Grounding refers to establishing the time, location, or actuality of a situation according to some reference point in the conversation~\cite{Clark1991GroundingIC}. Two people in a conversation need to coordinate not only the content of what they say but also how that message is delivered. For example, if Mary wants to get Clara to join her for lunch at a particular restaurant, she cannot simply email her with - ``Let’s meet at Sol at noon.''
After sending her invitation, Mary awaits evidence that Clara has received, understood, and committed to the lunch invitation. Meanwhile, Clara does not find a taxi as soon as she gets Mary’s message but sends an email response. To be at common ground, if Mary and Clara need to further clarify or modify their plans, they may exchange additional emails before they consider their plan to meet at the restaurant. 

With respect to dashboards, while onboarding (as discussed above) can assist in moving users through the grounding stage, there are other actions designers can take to build shared understanding of expectations and terms. The first might be the direct solicitation of priors and predictions from users. Hullman \& Gelman~\cite{hullman2021designing} suggest that existing (ungrounded) ``model free'' visualizations are inherently limited for visual analytics, and point to examples where either asking the user to predict data~\cite{kim2017explain} or, alternatively, showing users the predictions of others~\cite{kim2017draw}, can not only result in improved recall and retention of information, but also avoid drawing spurious conclusions. Shi et al.~\cite{shi2022fourth} similarly point to cases where data stories 
solicit information from users in order to ensure that the resulting information is \change{\textbf{relevant, interesting, or contextualized}} for users, and Lin et al.~\cite{lin-hunches} call for incorporating users' ``data hunches'' into charts.

\subsection{Turn-taking}
Turn-taking is a fundamental aspect of dialogue and occurs in a conversation when one person listens while the other person speaks~\cite{sacks:1974}. As the conversation progresses, the listener and speaker roles are exchanged back and forth. Participants need to coordinate who is currently speaking and when the next person can start to speak. Humans are very good at this coordination and typically achieve fluent turn-taking with very small gaps and little overlap. A conversationalist who does not allow others a turn, or speaks over others, may be considered rude. An example of turn-taking in conversation is:
\begin{dialogue}
\speak{Speaker A} ``Lovely weather this week.''
\speak{Speaker B} ``Isn’t it? I hope it's nice on the weekend.''
\speak{Speaker A} ``Me too. I have plans to go for a hike.''
\speak{Speaker B} ``That's fun! Which trail are you going on?''
\end{dialogue}


As dashboards move from static displays to more complex and interactive forms~\cite{sarikaya2018we}, there are an increasing number of examples of \change{\textbf{ bi-directional communication}} between the user and a visualization system. Examples of this communication can be as simple as providing tooltips or annotations on a user's request, supporting filtering or aggregation options, to more complex forms such as soliciting personal information from the user~\cite{shi2022fourth} or even incorporating ``analytical chatbots''~\cite{setlur:analyticalchatbots} that respond to natural language queries. Failure to allow the user to perform follow-up actions (as in \autoref{fig:teaser}) can result in frustrating analytical experiences where a user has a question or concern that the dashboard is not equipped to address. 
For example,~\cite{tory2021finding}  a sales dashboard that only allows a user to see a snapshot of the data at a single point in time can be frustrating if the user's next step is to try to understand the data in the context of the last month or year.

Dashboards systems can also take conversational initiative, and there are potential analytical benefits for such ``proactive design.''~\cite{tory2019mean} An example is the Frontier system~\cite{lee2022reccs}, where the user can select recommended views based on a set of analytical intents. Other forms of bidirectional interaction can be more subtle: for instance, the autocompletion metaphor in visual analytics~\cite{setlur2020pique} represents an attempt to match a user's utterance or intended action with the system's understanding of valid or popular alternatives. One consideration with turn-taking in dashboards is to allow bi-directional communication and useful division of labor between the person and the system, while respecting the user's agency and autonomy~\cite{heer2019agency}. Systems that steal focus, override user choices, \change{and lead to \textbf{dead-ends} in the communication flow,} are ``impolite''~\cite{whitworth2005polite} and produce friction and user enmity. 

\subsection{Repair and Refinement}
Conversational repair and refinement is the process conversation participants use to detect and resolve problems of speaking, hearing, and understanding~\cite{schegloff:1977}. If dialogue is to proceed smoothly, it is vital that there are opportunities for checking to understand and provide clarification when misunderstanding does occur. Everyday interaction is full of such checks and repairs, though these may be so automatic as to be almost seamless, rarely disturbing the flow of the interaction. In human conversation, there are continual implicit
acknowledgments that communication is proceeding smoothly. The speaker monitors the participants in the conversation in different ways to see if they understand (e.g., using checking moves such as ``Do you know what I mean?'') and the other participants are often giving verbal acknowledgments to the speaker (e.g., ``yes'', ``uh huh''). 
However, if the utterance is not understood, repair may be initiated. Through repair, participants display how they establish and maintain communication and mutual understanding during the turn-taking process.

Repair and refinement are both critical components of interactive dashboard design. NLIs for data provide a model for this sort of interaction, as natural language utterances (and the intents behind them) are often vague~\cite{Setlur2020SentifiersIV}, under-specified~\cite{setlur2019inferencing}, or misinterpreted by the natural language system. 
Some systems \change{\textbf{afford follow-up conversations}} for repairing or re-specifying intents. Perhaps more relevant to dashboard design are systems like DataTone's~\cite{datatone} ``ambiguity widgets'' that explicitly afford the resolution of ambiguous queries. The inability to update a dashboard when information is stale, irrelevant, incorrect, or misaligned with the user's goal can lead to frustration, as in \autoref{fig:teaser}.

Another way to support repair in analytic conversations with dashboards is to support fluid switching of tools and contexts if the existing dashboard is insufficient for a particular analytical task. Both Tory et al.~\cite{tory2021finding} and Bartram et al.~\cite{bartram2021untidy}, in their interviews with ``data workers'': reveal a recurring need to move data between tools (for instance, into a spreadsheet tool for manual data cleaning or inspection, or into a presentation tool for curated storytelling), and frustration with existing dashboard software that makes this process difficult.

A last intriguing potential for repair in dashboard design is to \change{\textbf{employ summaries or recommendations to prevent or ameliorate cognitive biases}} on the part of the users~\cite{wall2017bias,wall2019bias}. That is, a belief that the user is making a potential analytical error or oversight and intervening. For instance, Wall et al.~\cite{wall2022bias} propose the incorporation of a user's interaction records to provide a summary report explaining whether they are interacting with biased subsamples of the whole dataset, or whether they have considered representative facets of the data.

\subsection{Close}
Close is the process by which two partners end a conversation by offering and accepting each other’s final bids to close the conversation. Politeness strategies 
can avoid miscommunication when terminating the conversation. Coppock~\cite{coppock:2005} proposed several strategies
used to close the conversation: positive comment, excuse, and the imperative (e.g., ``it
looks like our time is up''). A positive comment implies that the conversation is pleasant, but the other does not want to continue. Excuse expresses an intent to end the conversation by providing an alternative motivation (e.g., ``I better get back to work''). The imperative strategy explicitly employs an imperative tone to end the conversation (e.g., ``It was nice talking to you'').

While the end of a specific analytical \textit{session} may be clear cut (say, navigating away from a website or closing a piece of software), a user's analytical \textit{conversation} does not end when they stop looking at a dashboard; the notion of a final close is more fraught. As users of dashboards are commonly impacted by \textit{reliance} on others~\cite{tory2021finding} (either for data, stakeholder buy-in, or discussion of goals), there is often a step of sharing the insights gleaned from an analytical conversation with various levels of formality and practice~\cite{brehmer2022jam}. \change{\textbf{Providing useful summaries of information or insights in a dashboard}}, and particularly summaries that can ``travel'' across different modalities, is, therefore, a critical (but often overlooked) component of dashboard design. Of particular interest to us is how summaries can concisely present not only the insights gained by the user over the course of an analytical conversation but also the supporting evidence for these insights (and the strength of this evidence).

Beyond post hoc summaries, we point to two potential examples of visualizations making good use of the \textit{end} of analytical conversations. The first involves systems where past users can provide important context for future users, as with Feng et al.'s~\cite{feng2017hindsight} Hindsight system where the interaction history of other users can be used to suggest potential starting places for new users, or in Kim et al.~\cite{kim2017draw} where other viewers' predictions can help situate one's own expectations of the relationships between data values. The second example embraces the multiplicity of potential methods and the potential fragility of conclusions, as in Dragicevic et al.'s~\cite{dragicevic2019multiverse} multiverse analysis reports, where the goal is to produce a report (with included conclusion and discussion sections) that is robust across a variety of different analytical choices or even natural data variability.

\begin{figure}[t!]

    \centering
    \includegraphics[width=\columnwidth]{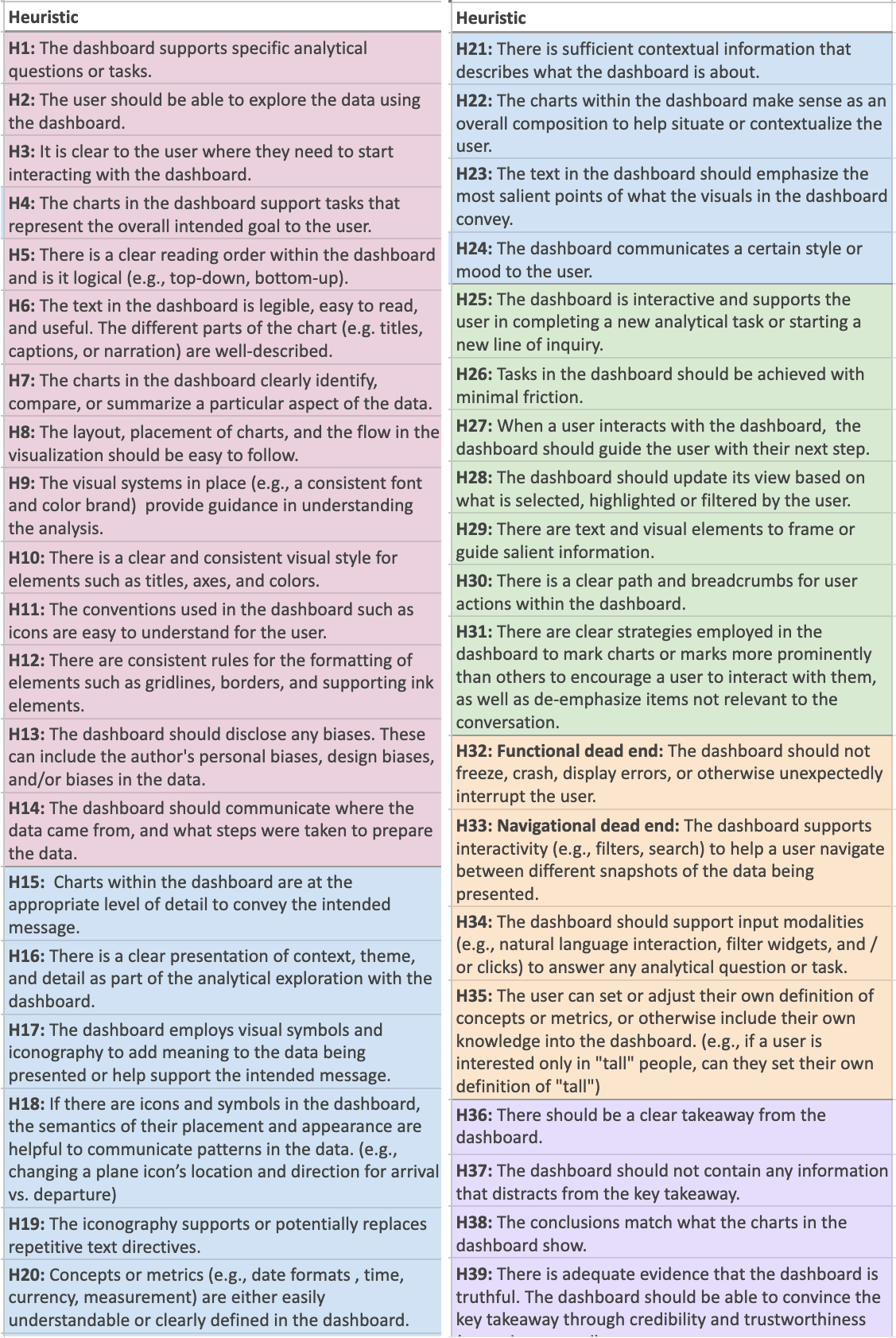}

  \captionof{table}{Through an iterative process, we distilled $39$ heuristics for analytical conversation organized into the five conversation states: \textcolor{initiationcolor}{\textbf{initiation}}, \textcolor{groundingcolor}{\textbf{grounding}}, \textcolor{turntakingcolor}{\textbf{turn-taking}},
   \textcolor{repaircolor}{\textbf{repair \& refinement}}, and \textcolor{closecolor}{\textbf{close}}.\change{A more detailed version of this table with the labeled categories and assigned Gricean Maxims is provided in supplementary material.}}
      \label{tab:heuristics}
\end{figure}

\section{Iterative Development of Conversation Heuristics}
We apply the notion of cooperative conversation and its maxims by examining the conversational properties that are specifically relevant to interactive dashboards, drawing from the following sources: 

\begin{itemize}
\item \textbf{Natural language interfaces for visual analysis}: We explore how language pragmatics in the context of natural language interfaces can help support analytical conversation. A review of previous academic prior art and software systems that implement techniques for supporting analytical conversation in the context of NLIs for visual analysis~\cite{datatone,setlur2016eviza, hoque2017applying,setlur:analyticalchatbots,srinivasan2017orko,powerbi,ibmwatson,thoughtspot} provided guidelines for informing the various heuristics for supporting the various conversational states when interacting with data.
\item \textbf{Cooperative conversation behaviors in human-computer interfaces}: The design of such interfaces often draws inspiration from human-to-human conversation and mechanisms that facilitate the exchange of information between speaker and listener. There exists an expectation that the information shared is relevant and that intentions are clearly conveyed to support a cooperative conversation that is truthful, relevant, concise, and clear. A review of the various applications of Gricean Maxims and cooperative conversation guidelines in interactive interfaces and experiences between humans and computers, ranging from human-bot interaction, chatbots, smart assistants, and embodied agents~\cite{Cassell2001EmbodiedCA,En2011TheAO,panfili:2021} helped define the various heuristics that satisfy the maxims.
\item \textbf{Practioner examples of dashboard design}: Interactive design guidelines for authoring functionally useful interactive dashboards as described in practitioner literature~\cite{knaflic2015storytelling,setlur2022functional}, provided examples to help inform the creation of the initial set of heuristics.
\end{itemize}

However, as indicated in Section~\ref{sec:relatedwork}, many of the guidelines from the visualization literature tend to focus on recommendations and best practices for layout, visual composition, data encodings, and chart types, as well as for natural language interfaces and systems. We instead focus on Grice's Cooperative Principle and its associated maxims as a way to identify themes to support analytical conversations in interactive dashboards. In particular, we apply the notion of \emph{conversational implicature} as a way to systematize the properties of interactive dashboards. Conversational implicature, as introduced by Grice, is an indirect or implicit act within a conversation, determined by the conversational context that supports the primary dialogue~\cite{Grice1975-GRILA-2,sep-implicature}. Implicature serves a variety of conversation goals towards effective communication, supporting pragmatics, maintaining good social relations, and overall efficiency in conveying the intended message. \change{To come up with an initial set of heuristics, the co-authors adapted guidelines and heuristics developed for natural language interfaces to interactive dashboards (e.g., ``Does the dashboard freeze, crash, display errors, or otherwise unexpectedly interrupt the user?'') and drew inspiration from example dashboards authored by visualization experts (e.g., ``Is there a clear reading order and is it logical (e.g., top-down, bottom-up)?''.}


All the co-authors iteratively developed a set of 
heuristics, organized into themes, that support conversational implicature through \emph{both} the presentation and interaction of dashboards with a human. Each co-author picked one of three dashboard examples \change{of their choice that they encountered recently} - \change{a Tableau Public dashboard showing the best states to retire in the US~\cite{USStatesRetireDashboard}, a COVID-19 Dashboard~\cite{WACovidDashboard}, and a Tableau World Indicators Business Dashboard~\cite{WorldIndicatorsDashboard}} and independently reviewed the current heuristics to assess if they were relevant \change{(including whether they were supported or violated)} to the corresponding dashboard example. Subsequently, the co-authors collectively discussed and compared insights on what it meant for a dashboard to be cooperative.

We initially collected  $95$ potential heuristics. Note that we chose the term `heuristic' defined as
``serving as an aid to learning, discovery, or problem-solving by experimental and especially trial-and-error methods'' and ``relating to exploratory problem-solving techniques that utilize self-educating techniques to improve performance''~\cite{mw:heuristic} as a means to help guide a dashboard author. Through our experience in using the heuristics, we clustered them into related themes and iteratively reworded and clarified them to minimize unfamiliar jargon or other vague terms. This process resulted in $56$ heuristics. 

\subsection{Phase 1: Pilot Review}
\label{sec:heuristics}
We tested the modified set of heuristics with two pilot participants. The instructions asked each participant to pick an interactive dashboard that they recently authored, run the heuristic checklist by the dashboard, and respond with detail about whether the dashboard supported the given heuristic or not. Lastly, they were asked to indicate if any of the heuristics were confusing to understand or apply. Based on feedback from this exercise, we updated the instructions to include an example along with a screenshot of a sample dashboard for a heuristic, refined and consolidated the heuristics further, resulting in $53$ heuristics under $13$ themes \change{(analytical conversation support, multi-modal conversation support, use of semiotics, clarification of vague concepts, communication goal, summaries and takeaways, exposition, integrating text with visual information, composition and layout, visual scaffolding, level of detail, trust and transparency, register)}. 



\subsection{Phase 2: Expert Feedback}
\label{sec:expertfeedback}

We then conducted a self-reflection exercise with 16 expert visualization researchers and practitioners. \change{Based on self-reporting, experts comprised six business intelligence analysts, five data visualization consultants, and five Ph.D. visualization students with at least three years of experience authoring visualizations and dashboards.} One participant did not complete the exercise, leading to a total of 15 completed exercises. The goals of Phase 2 were to 1) understand how the heuristics are applied when critiquing the design and interaction of a dashboard and 2) get feedback about the clarity and usefulness of the heuristics. 

\begin{figure*}[ht]
    \centering
\includegraphics[width=.9\textwidth]{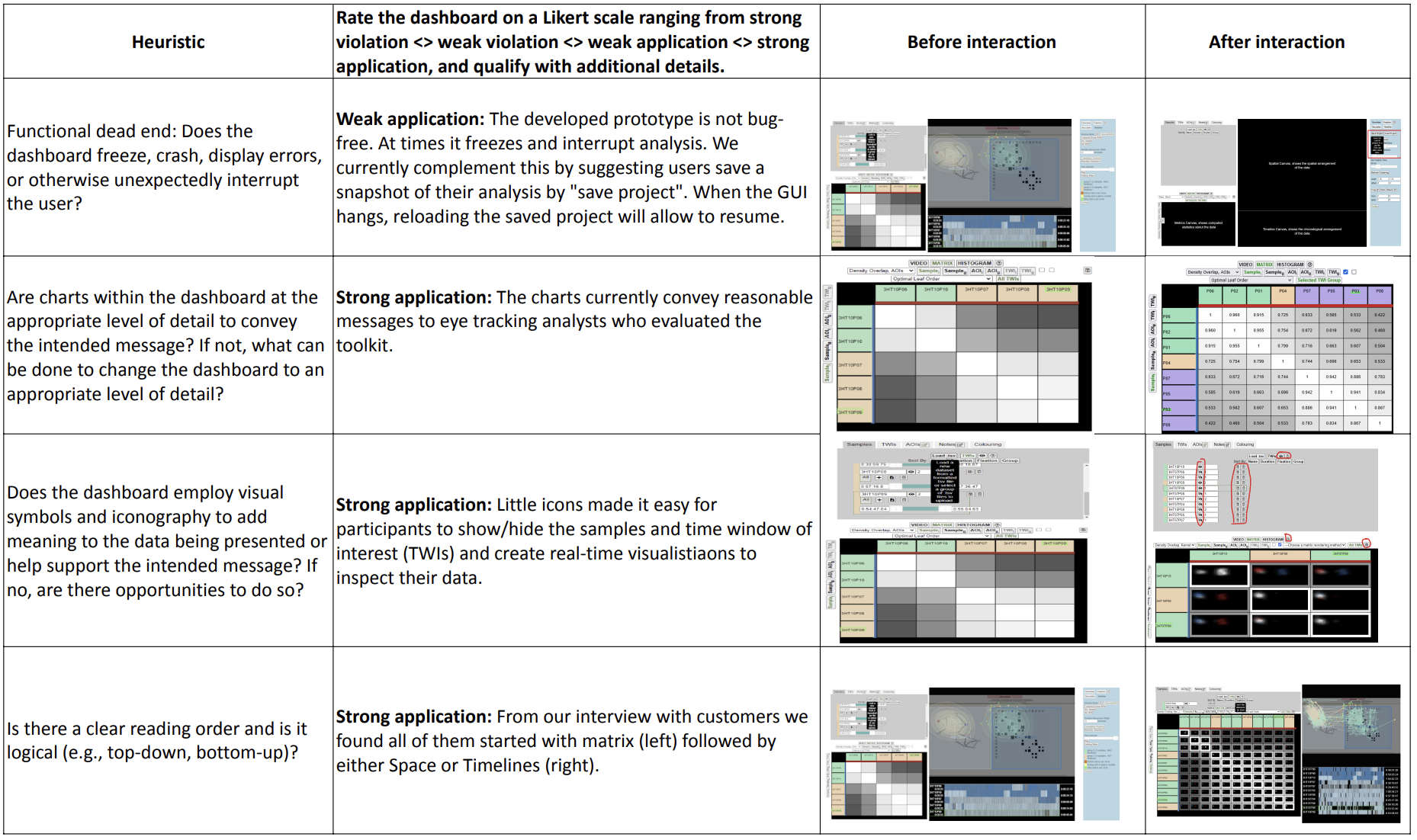}
  \caption{\change{Example of four heuristics used to evaluate a visual eye tracking dashboard by an expert. The table indicates the heuristic, its violation or applicability to the dashboard, and the states of the dashboard before and after the interaction, respectively.} }
      \label{fig:expertreflection}
\end{figure*}

\subsubsection{Expert Reflection Exercise}
\label{sec:reflection}
The user study was designed as a self-reflection exercise where participants were asked to evaluate each heuristic against a dashboard example that did not contain confidential or proprietary data. We asked them to include a link and a screenshot of the dashboard they picked with an explanation for their choice. We included a link to a spreadsheet of heuristics, and for each heuristic, the spreadsheet asked participants to first determine if the heuristic applied to their chosen dashboard and, if not, to explain the reason. They were also asked to rate the extent of the application or violation on a 5-point semantic differential scale ranging from ``Strong violation'' to ``Strong application.'' The spreadsheet also requested participants to provide visual examples of applications and violations of the chosen dashboard for each heuristic wherever possible. To help the participants understand the expectations for the exercise, we provided an example response to one of the heuristics.

After the participants completed the heuristics spreadsheet, they were requested to answer a set of questions:
\begin{itemize}
\item Were the heuristics and/or themes useful? How? Which ones in particular? Explain in detail.
\item Were any of the heuristics not helpful or confusing to you? If so, please elaborate.
\item Did any of the heuristics make you think of dashboard design in a new way?
\item Were there any heuristics that you thought were missing?
\item What changes would you make to your dashboard based on this assessment? Please describe in detail.
\item Do you plan on updating your dashboard in response to these heuristics? If so, would you be willing to send us an update?
\end{itemize}

We estimated the study would take approximately $45$ minutes to complete. Participants were given three days to complete the study on their own time and were compensated with a $\$30$ Amazon gift card. We recruited the expert participants (indicate by the notation [$E\#$]) through a screening survey (included in supplementary material) posted on social media channels and distribution lists at a large software company. Participants were required to have experience (at least five years) designing or evaluating interactive dashboards using software like Tableau or PowerBI, notebook environments like Jupyter or Observable, or libraries like D3 or matplotlib. We also required participants to have a dashboard they were working on and that they were willing to share with us in some form (as a web link or a screenshot). We collected background information of the survey respondents that included a description of their current job role, years of experience designing dashboards, and a description of the topic and the audience of the interactive dashboard that they were designed for.

\subsubsection{Assessing the Utility of the Heuristics}
To assess the utility and comprehensibility of the heuristics, we reviewed participant responses for the following scenarios:
\begin{tight_itemize}
\item \textbf{Heuristics indicated as `does not apply'}. Instances where participants indicated that the heuristic was not relevant to the dashboard they were evaluating.
\item \textbf{Misinterpreted or hard to understand heuristics}. Instances where participants misinterpreted a heuristic for another or simply did not understand them.
\item \textbf{Heuristics marked as `strong violation' / `weak violation'}. Instances where participants indicated that their dashboard violated a given heuristic.
\item \textbf{Heuristics marked as `strong application' / `weak application'}. Instances where participants indicated that their dashboard satisfied a given heuristic.
\item \textbf{Duplicate or similar heuristics}. Instances where participants marked two or more heuristics as either duplicates or very similar. 
\end{tight_itemize}

\subsubsection{Expert Responses}

All co-authors inspected the 15 expert responses. The expert participants chose dashboards that they had authored for an audience that included either a client, a data visualization class, or sharing on Tableau Public. The themes of dashboards ranged from health monitoring, crime and violence, visual eye tracking analytics, sports, and finance. Figure~\ref{fig:expertreflection} shows an example dashboard assessed by the heuristics. Here is an overall summary of how the heuristics were labeled:
\begin{tight_itemize}
\item \textbf{Heuristics indicated as `does not apply'}. $23$ of the $53$ heuristics were labeled as ``does not apply'' to their dashboards by at least one participant. For example, several participants marked the heuristic, \textit{``Does iconography support or potentially replace repetitive text directives?  If not, are there opportunities to do so?''} to be not applicable to the dashboards they were assessing.
\item \textbf{Misinterpreted heuristics}. $28$ of the heuristics were marked as difficult to interpret by at least one participant. For example, participants reported having trouble understanding heuristics that were rather vague: \textit{``Does the dashboard support open-ended data exploration? If not, why?''} or contained jargon: \textit{``Does the visualization disclose the provenance of the data?''}

\item \textbf{Heuristics marked as ` strong violation' / `weak violation'}. 
\change{On average, $11$ out of $53$ heuristics were marked as being either strongly violated or weakly violated (min: $0$, max: $27$).} For example, for the heuristic, \textit{``Are vague concepts clarified if they exist within the data? (e.g., tall or high-performing) If no, which vague concepts should be clarified?''} was commonly marked as a `strong violation'. $E8$ commented, \textit{``This dashboard is meant for the public but uses many difficult terms like `Case trajectory' instead of ``number of people with covid'' and `wastewater concentration.' We should put the text through a plain language grader and improve the language for ease of understanding.''}
\item \textbf{Heuristics marked as `strong application' / `weak application'}. 
\change{On average, $39$ out of $53$ heuristics were marked as being either strongly applicable or weakly applicable (min: $21$, max: $49$).} We hypothesize that given that the dashboards are authored by experts, a high number of heuristics were labeled as applicable to the dashboards. For instance, most participants ($13$ out of $15$) stated that \textit{``Is the dashboard interactive to support the user in completing a new analytical task or starting a new line of inquiry? Are there interactions that could be added to enhance the experience?''} strongly applied to their dashboards. $E11$ marked, \textit{``If there is interaction, does the dashboard update as expected?''} as a `weak application' and commented  \textit{``Filtering and hovering interactions update the story as expected. But the lack of instructions makes the user perceive filters as labels.''}
\item \textbf{Duplicate or similar heuristics}. Six sets of heuristics were marked as either being duplicates of one or more other heuristics or very similar. For example, under the theme, ``Composition, layout, space, and sequencing'', heuristics such as \textit{``The layout, placement of charts, and the flow in the visualization should be easy to follow''}, \textit{``There is a clear reading order within the dashboard and is it logical (e.g., top-down, bottom-up)''}, and \textit{``The charts, text, and any other visuals are laid out in a way that is helpful for understanding the structure of the information being presented in the dashboard''} were identified to be similar.
\end{tight_itemize}

Generally, participants found the dashboard reflection exercise to be helpful. $E14$ said, ``\textit{ The heuristics and themes are very helpful in understanding many of the considerations that need to be made while designing a dashboard such as (1) Multi-modal conversational support, (2) Integrating text with visual information for communication, and (3) Visual scaffolding for helping with conversation clarity.}'' Participants also found that the reflection inspired them to consider dashboard design in new ways. $E05$ said, ``
\textit{Multimodal interactivity and NLI provide a new way of thinking. It would be exciting to integrate this in an eye-tracking analysis tool for improving the sense-making loop.}'' 

After reviewing the experts' reflections, we clarified heuristics that were unclear and ambiguous as well as consolidated redundant ones, resulting in a total of $46$ heuristics. For example, we removed redundant heuristics such as \textit{``The quantitative units are clearly defined or specified.''} as we already included the heuristic, \textit{``Concepts or metrics are either easily understandable or clearly defined in the dashboard.''} and added heuristics suggested by the participants, such as \textit{``Is there adequate evidence that the dashboard is truthful? Is the dashboard able to convince the key takeaway through credibility and trustworthiness?''}

\subsection{Phase 3: Author Reflections and Final Iterations}
We further reflected on the set of heuristics, given that the goal was to evaluate them with a student population to assess how the students would apply and critique the conversational nature of dashboards. We reformulated the remaining heuristics to follow a clear and consistent format and to clarify issues identified by the expert evaluators. Specifically, we further iterated on the heuristics based on the criteria:
\begin{tight_itemize}
\item Reworded heuristics posed as questions to be imperative guidelines of what the dashboard ought to support. For example, heuristic \textit{``Is the text in the dashboard legible, easy to read, and useful? Are the different parts of the chart (e.g., titles, captions, or narration) well-described?''} was rephrased as \textit{``There are text and visual elements to frame or guide salient information.''}
\item Ensured that the heuristics were understandable without technical jargon where heuristics such as \textit{``Do starting points for interactivity align with user experience and expectations?''} were reworded as \textit{``The dashboard is interactive and supports the user in completing a new analytical task or starting a new line of inquiry.''}
\item Made sure that each heuristic could be clearly validated for whether it was applied or violated in an interactive dashboard. To that end, any conjunctions, if present, were removed to prevent the inclusion of multiple guidelines within a single heuristic.
\end{tight_itemize}

Finally, after winnowing down the heuristics to $39$, we found that they could be reorganized thematically into the $5$ basic conversational states: initiation, grounding, turn-taking, repair \& refinement, and close. While the final set of heuristics provides an initial framework for assessing cooperative conversation in interactive dashboards, we do not guarantee completeness; rather, we sought to assess their utility and identify opportunities to further improve and refine them. The next section describes how the heuristics were applied by students in a visualization education setting. 
The final table of conversational dashboard heuristics is shown in Table~\ref{tab:heuristics}, and its various iterations leading to the final set are included in the supplementary material.

\section{Use of Heuristics in Education Practice}

To evaluate the utility of the heuristics, we provided two opt-in homework exercises with visualization learners in a post-graduate data visualization class at a university. Part A was a heuristics reflection exercise on pre-authored interactive dashboards, while Part B was an exercise to apply the heuristics to improve the conversational nature of an existing dashboard. Both exercises were not graded to mitigate any biases when students provided feedback. The university review board granted formal approval to conduct the exercises. We include class exercise material and evaluations as supplementary material.

\subsection{Part A: Heuristics Reflection Exercise}
The goals of the heuristics reflection exercise were to 1) assess the heuristics' value in supporting visualization learners and 2) gain feedback on the heuristics for iterative improvement. Since our development phases involved visualization experts, we focused on learners to ensure the heuristics were understandable by a less experienced population. 

The $52$ participants were master's students (with backgrounds in Computer Science or Engineering). We use the notation [P\#] when referring to participants in this heuristics evaluation. We refer to particular heuristics from our final list as [H\#]. 

The homework exercise was conducted similarly to the reflection exercise described in section~\ref{sec:reflection} but with the updated heuristics table (Table~\ref{tab:heuristics}, organized by the five conversational states). The exercise was introduced during class by the class instructor and then completed as a homework assignment over a week. \change{To ensure that students remained engaged when applying the heuristics to evaluate dashboards, we provided a list of 18 dashboards and asked students to describe which dashboard they chose and why. Four dashboards were not picked from the list, with the highest number of students (six) choosing a renewable energy consumption dashboard. The complete list of dashboards and the frequency of choices is included in the supplementary material.} 
The actual reflection exercise was the same as in section~\ref{sec:reflection}; it involved assigning the dashboard a rating (`strong application,' `weak application,' `weak violation,' or `strong violation') for each heuristic with written commentary and screenshots to justify the ratings and then answering the reflection questions. The students additionally gave an in-class presentation of their findings from the homework exercise.

We conducted a thematic analysis of the heuristic reflection responses and survey answers. We looked for feedback on the heuristics, interesting examples of how the heuristics were applied to the dashboards, and insights that were revealed. We also examined frequency data on how dashboards were ranked across the different heuristics. 
\vspace{-1mm}
\subsubsection{Rating Frequencies}
\begin{figure}[ht]
\vspace{-3mm}
  \centering
  \includegraphics[width=.6\columnwidth]{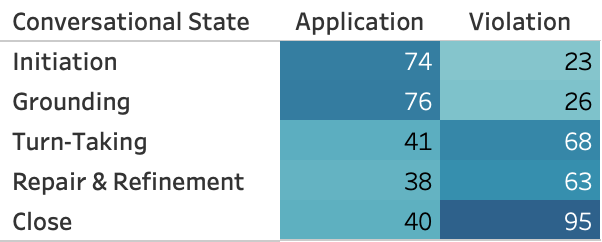}
  \captionof{table}{\label{fig:heatmap}
           \change{Part A }heatmap showing the frequency of dashboards applying or violating the heuristics, grouped by conversational state. Frequencies are averaged across participants and normalized by the number of heuristics per conversational state on a 0 to 100 scale.}
\end{figure}

\begin{figure*}[ht]
  \centering
  \includegraphics[width=3.6in]{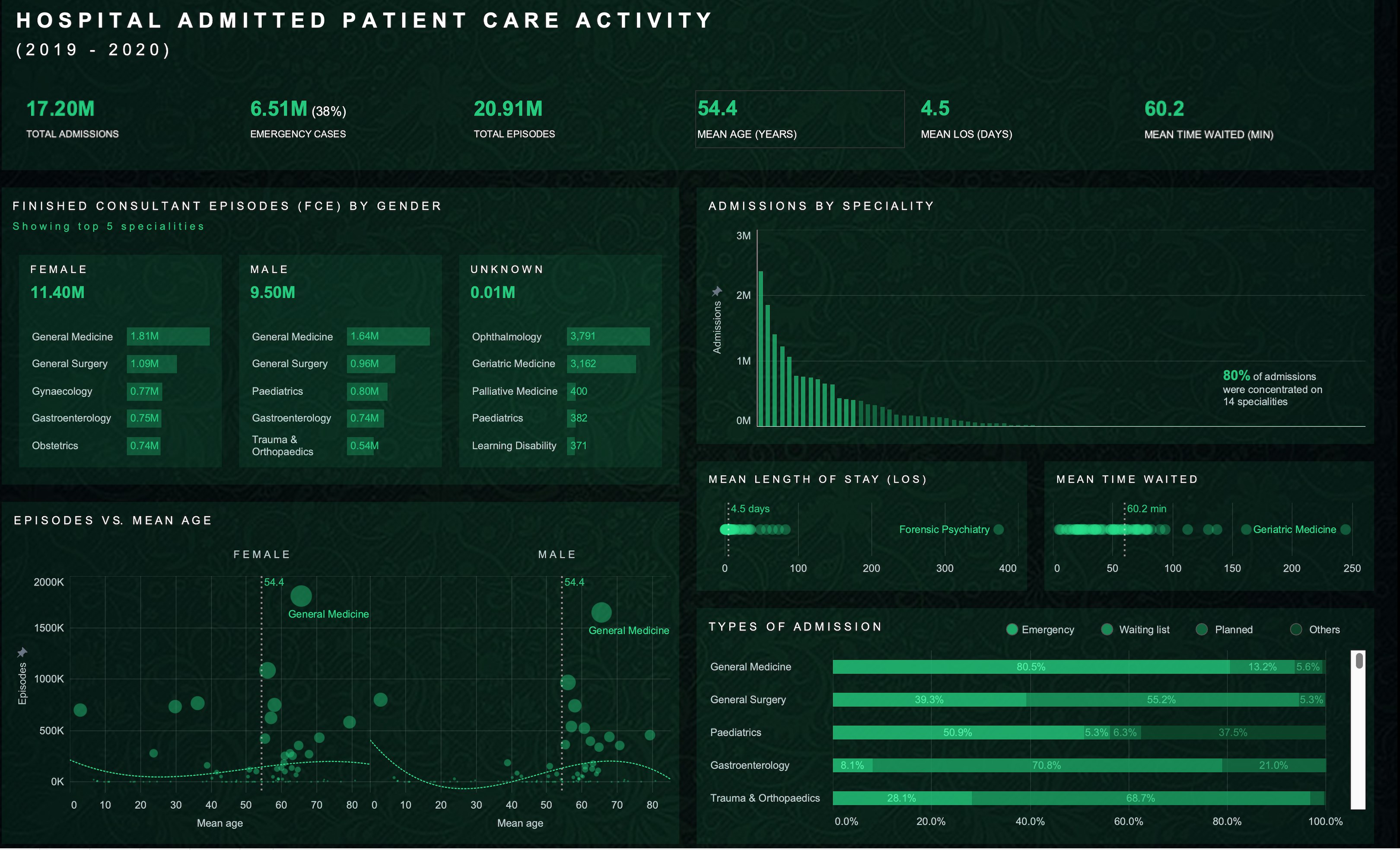}
    \includegraphics[width=3.4in]{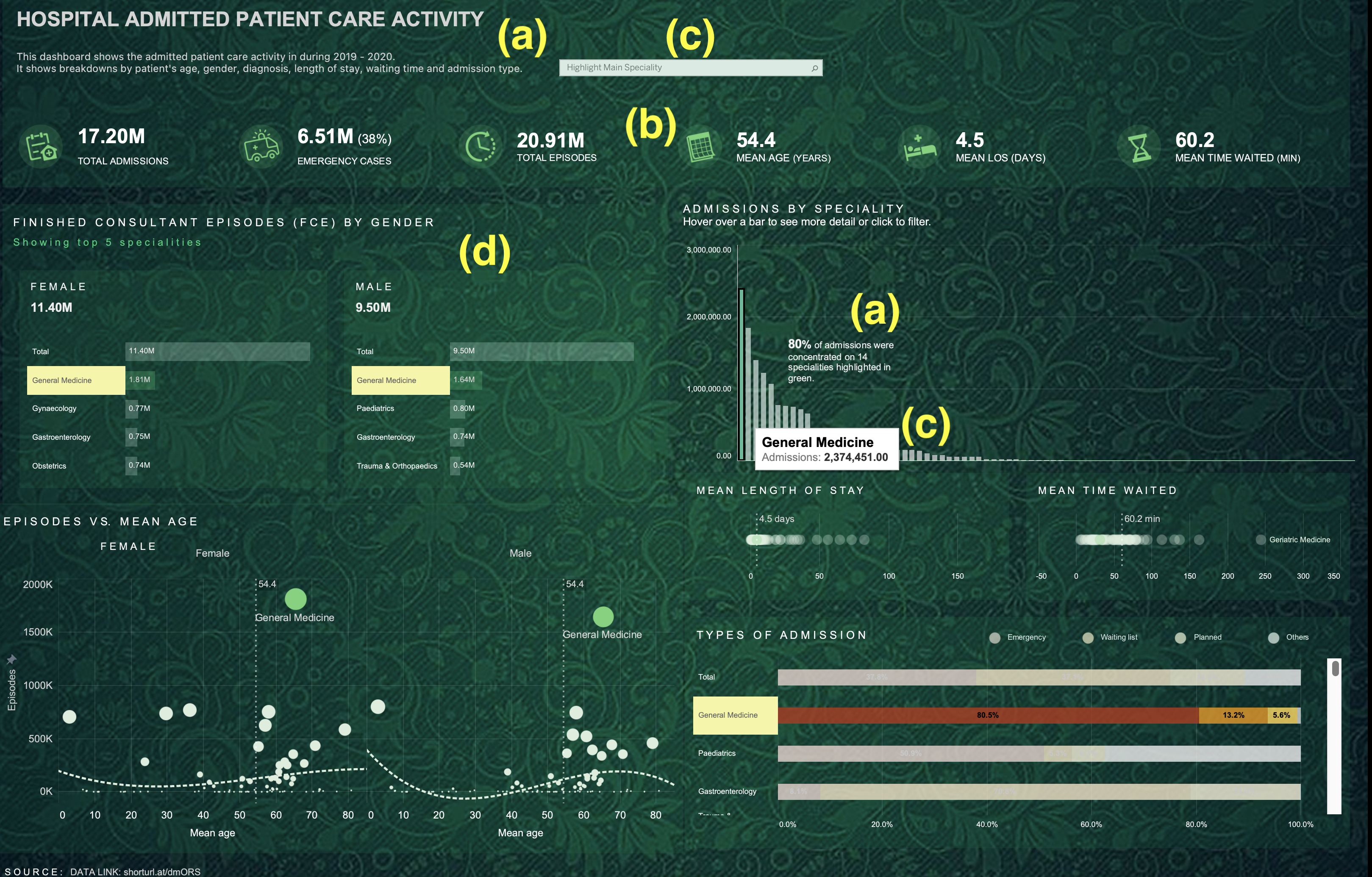}
  \caption{\label{fig:planB} Dashboard used in Part B classroom exercise. Left: Original dashboard. Right: Modified dashboard. Updated dashboard annotated in yellow based on heuristics from the conversational states, indicated by $H\#$: (a) Initiation: The text in the dashboard is updated to provide more context ($H6$). (b) Grounding: Dashboard contains iconography to add meaning to the data being presented ($H17$). (c) Turn-taking:  The dashboard is updated to add filtering across views ($H28$). (d) Close:  The "Unknown" category was removed from the original dashboard to further clarify the takeaway of the dashboard ($H37$). \change{Note that students made aesthetic changes (not always an improvement), such as modifying the dashboard's background in addition to applying the heuristics.} (Permission granted to use original and modified dashboards).}
\end{figure*}

Relative frequencies of heuristic applications and violations, as rated by participants for their chosen dashboard, are summarized in Table~\ref{fig:heatmap}. Because the conversational state categories contain different numbers of heuristics, we used a normalized metric rather than raw counts. To compute these scores, we first combined strong and weak application ratings, and similarly combined strong and weak violation ratings. We averaged the number of ratings across participants and normalized the result by the number of heuristics in each state on a $0-100$ scale. Note that these are not exactly percentages because a participant could identify multiple applications and/or violations of a single heuristic.

Table~\ref{fig:heatmap} shows that the rate of violations increased for later conversation phases, and the rate of applications decreased. This observation was consistent across participants. It suggests that today's dashboards offer reasonable support for initiation and grounding, but are progressively less supportive as human-data conversations get into turn-taking, repair, and close activities. For instance, turn-taking repeatedly showed up as a challenge, where dashboard inflexibility or awkward interactions made it difficult for users to complete analytical workflows. 

\subsubsection{Use of Heuristics Across Conversational States}
\label{sec:partAScores}
Next, we examine themes and interesting examples of how the heuristics were used across the conversational states.

\pheading{Initiation}. Heuristics in the initiation state ($H1-H14$) were often marked as either strong or weak application (normalized application frequency of $74$ in Table~\ref{fig:heatmap}). Participants noted that dashboards initiated the conversation by including instructions on how to use the dashboard, making it easier to explore the data. The reading order, encodings, and formatting conventions used were often easy to understand and follow. $P13$ stated \textit{``the color combination used by the chart maker keeps the reader attentive and focuses the attention at the right regions.''} However, dashboards did have violations in revealing the provenance of their data. $P8$ stated, \textit{``strong violation as the dataset source hyperlink they tried to give doesn't work and the data preparation is not mentioned.''}

\pheading{Grounding}. Similar to the initiation state, heuristics in this conversation state ($H15-H24$) were often marked as either strong or weak application (normalized application score of $76$). Many of the dashboards ($38$ out of $52$) were described as having a clear presentation of context and level of detail. $P38$ commented, \textit{``Yes ordering is logical, It's sorted in highest to lowest expense. First row shows line chart and next row shows details of breakdown.''}

\pheading{Turn-taking}. For this conversational state, there was a lower frequency of application ratings ($41$) and a higher frequency of violations ($68$), indicating the limited interactivity ($H25$) that the dashboards provided. $P20$ stated, \textit{``The dashboard should update its view based on what is selected, highlighted, or filtered by the user. As there are no filters and update options available in the dashboard.''}
Participants also noticed some friction when interacting with the dashboard ($H26$). $P3$: ``\textit{The process of zooming in is clunky and disrupts continuity.}'' and guiding to the next step. $P3$: ``\textit{Very little visual warning/cueing to accompany changes to graphs, particularly in the side panel. Some changes are initially off screen and have to be scrolled to.}''

\pheading{Repair \& Refinement}. Participants ($46$ out of the $52$ students) often found that the dashboards violated the functional and navigational dead-ends ($H32$ and $H33$) ($63$ violations per 100 cases). $P28$ commented, \textit{``The dashboard doesn't provide interactivity at all. It has no filters or searches. Just a basic static visual. Just looking at the graph doesn't make any sense unless we hover over it.''} Further multi-modal support ($H34$) was violated in many cases as the interactions were limited to selecting filters in the drop-down, for example. $P41$ said, \textit{``There are no filters. Filters could have helped a lot when analyzing certain time periods but are given as only two values between year ranges.''}

\pheading{Close}. This category had the highest frequency of violations ($95$ violations per 100 cases). For several dashboards, it was not apparent what the key takeaway was to close the conversation ($H36-H38$). $P36$ commented, \textit{``Weak Violation. Just by looking at this dashboard, one cannot conclude something; the user has to gather data from each hexagon, then analyse it and only then something can be concluded.''}
Other violations concerned around trust ($H39$). $P5$ said, ``\textit{Strong violation: though there's no reason the believe the dashboard is \emph{lying}, without key context a user with no additional information could easily come away with a confused message, or even the wrong idea entirely.}'' Similarly $P45$ said, \textit{``Weak Violation. The source of the data is nowhere mentioned, which would have increased the credibility of the dashboard.''}

In summary, we found that while dashboards tend to be effective at initiation and grounding
of the conversation, they struggle with other aspects of conversation that include turn-taking, repair \& refinement.

\subsection{Part B: Update or Create Dashboards Using Heuristics}

The $52$ students self-organized into groups of three or four, forming a total of $15$ groups where they applied the heuristics to update an existing dashboard from Tableau Public (4 out of 15 groups) or create new dashboards from a Kaggle dataset~\cite{kaggle} (11 out of 15 groups). Students completed the exercise over a week and rated the dashboard with the same set of heuristics (Table~\ref{tab:heuristics}), providing commentary and screenshots. For exercises involving updating an existing dashboard, students rated the dashboard before and after the update.

We conducted a thematic analysis of the heuristic reflection responses. We looked for feedback on the heuristics, interesting examples of how the heuristics were applied to the dashboards, and insights that were revealed. We also examined frequency data on how dashboards were ranked across the different heuristics.

\subsubsection{Application of Heuristics Across Conversational States}

\begin{figure}[ht]
\vspace{-3mm}
   \centering
    \includegraphics[width=0.5\textwidth]{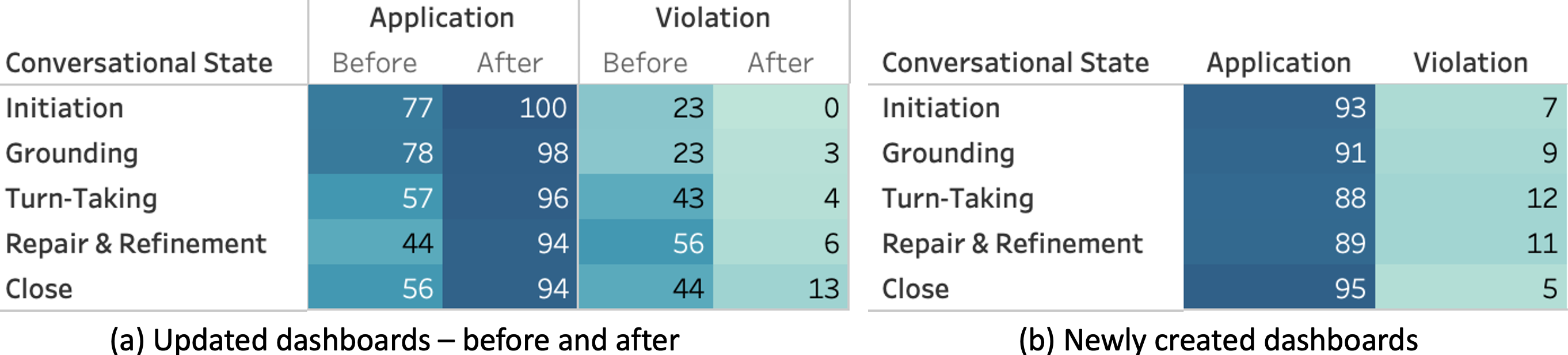}
     \captionof{table}{\label{fig:heatmapPartB}{\change{Part B }heatmaps showing the frequency of applying or violating the conversational heuristics, grouped by conversational state. Frequencies are averaged across participants \change{and normalized by the number of heuristics per state on a 0 to 100 scale.}}}
\end{figure}

Similar to Part A (Section~\ref{sec:partAScores}), we computed frequencies of heuristic applications and violations, as reported by students (Table~\ref{fig:heatmapPartB}). In cases where students modified an existing dashboard, students reported an overall increase in the rate of applications and a decrease in the rate of violations of heuristics across all conversation states. In particular, we saw a higher rate of decrease in violations for `turn taking,' `repair \& refinement,' and `close'; states that had fewer application rates in general during Part A's exercise. However, note that students performed these self-evaluations, which could contribute to a higher rate of applications. While the class instructor reviewed the self-reflection ratings, future work should consider an external reviewer to validate these ratings. Figure~\ref{fig:planB} shows an example of an original dashboard on hospital admittances (left) with a corresponding modified version (right). Updates include supporting better turn-taking by adding interactivity and multi-view coordination, along with a search bar to navigate to a specific medical department specialty. Iconography was added to better convey the meaning of the information being presented to the user, along with additional descriptive text to ground the conversation.

We found that having done Part A, students were familiar with the heuristics and focused specifically on addressing heuristics for turn-taking, repair \& refinement, and close. $P18$ stated, \textit{``I was more observant of how the dashboard behaved when I interacted with it. I focused on making sure there were no dead-ends when I clicked on the widgets and all the views updated appropriately.''} In both the updated and newly created dashboards, we observed a greater prevalence of text to help ground contextual information alongside the visualizations (applying heuristics, Initiation - H6, Grounding - H16, H21, H23, Turn-taking - H29, and Close - H36). $P4$ remarked, \textit{``For each category, the heuristics reminded me that text plays a vital role with the charts for communicating the key ideas.''} Some students reported that they sacrificed visual style for clearer communication with the user. $P37$ stated, \textit{``Although visual style get [sic] little disturbed in the color part, it looks necessary to make dashboard more easy to understand.''} Future work should further explore how these heuristics, alongside visual design guidelines can support the dashboard authoring process.


\subsection{Feedback on the Heuristics}
Now, we summarize the various qualitative themes of feedback on the heuristics across both exercises.

\pheading{Heuristics were helpful and understandable.} Participants found the heuristics to be useful for understanding the structure and flow of dashboards as part of an analytical conversation, as well as for authoring new ones. $P8$ commented, \textit{``The dashboard communicates a certain style or mood to the user, and there are clear strategies employed in the dashboard to mark charts or marks more prominently to encourage a user to interact with them, as well as de-emphasize items not relevant to the conversation.''}. $P4$ stated, \textit{``all heuristics were explained clearly, and I did not encounter any confusion while completing the form.''}

\pheading{Unique and unexpected heuristics.} Others found some heuristics to be rather unique and unexpected when considering dashboard design. For example, participants found heuristics, \textbf{$H11$, $H17-19$} on visual symbols and iconography to be helpful - \textit{``The use of semiotics for symbolic communication as well as the exposition sections stood out to me in particular [$P4$].} $P15$ found $H26$ to be useful when thinking of evaluating friction in dashboards: \textit{``I was not able to find or look at all the cities at once, and it was difficult to click on the small bubbles.''} $P13$ was intrigued by the heuristic on navigational dead-ends ($H33$) and said, \textit{``This made me think to make visualization work in every case, whenever the user selects or searches anything on the dashboard, to navigate easily.''} $P23$ found $H5$ concerning logical reading order to be insightful - \textit{``Before this assignment, I never thought that the placement of charts should have a logical sequence. It makes perfect sense, and I will apply this in my future dashboards.''}

\pheading{Confusing and missing heuristics.} Participants found the heuristics on bias ($H13$) and mood ($H24$) to be vague and not very actionable. $P11$ stated, \textit{``This [the bias] heuristic was confusing as I did not understand the biases in the dashboard very well. Mood communication in terms of the dashboard was somewhat confusing to me.''} There were suggestions for considering adding animation as part of the analytical conversation ($P24$, $P27$, $P36$) and further helping users recognize, diagnose, and recover from errors when interacting with the dashboards ($P47$).

\section{Discussion and Future Work}


While existing dashboard guidelines capture visual \change{design} issues such as legibility and complexity, our development and evaluation of heuristics from the lens of analytical conversations suggest that there are ways that dashboard design can succeed (or fail), which are not captured by existing recommendations or pedagogy, and so are often overlooked.

\pheading{Dashboards struggle with  turn-taking, repair \& refinement.} Participants 
pointed out many violations of heuristics in the turn-taking and repair \& refinement phases, 
suggesting that today's dashboards may be weak in these aspects. We strongly encourage future work that 
\change{makes dashboards} more flexible, cooperative conversational partners. Future dashboards could enable users to more easily pivot between analytical goals (e.g., \change{via} flexible construction so the end user can change dashboard metrics, field ordering, chart type, \change{etc.}) and could employ predictive analytics to anticipate a user's upcoming information needs.

\pheading{Interpreting heuristics for guidance and mitigation strategies.} \change{By their nature, heuristics offer guidance rather than prescriptive solutions. They should be considered in the context of the designer's expert knowledge of the domain, design goals, and audience. Heuristics may, at times, contradict each other or suggest design directions that are counter to specific communication goals or domain conventions. For example, in Sarikaya et al.'s~\cite{sarikaya2018we} framework of dashboard types, dashboards for learning may need greater emphasis on grounding (e.g., contextual information) than dashboards for ongoing awareness of well-understood metrics. We envision that designers will use the heuristics to inspire ideas and identify potential gaps and flaws while thoughtfully discarding less suitable suggestions. Utilizing the heuristics to provide in-situ mitigation strategies in dashboard authoring tools is an area of future research. For example, tools could flag warnings if the interactions have errors or there are no graceful fallbacks for preventing functional dead-ends. Other guidelines can support authors with progressive disclosure of content through interaction and templates for adding text to prevent cognitive overload during conversational initiation. }

\pheading{Developing heuristics for conversations around data.} We encourage the revision and extension of the heuristics themselves, as academics and practitioners use and adopt them. 
For example, we introduced heuristics to guide dashboard design and evaluation, with the lens of dashboards as a medium to enable conversations \textit{with} data. Dashboards also support the important role of human-human communication \textit{around} data~\cite{tory2021finding}, including discussing and circulating the data within an organization. A future extension to the
 \change{heuristics could 
focus on dashboard characteristics to support circulation or persuasion.}

\pheading{Assessing utility of heuristics during dashboard design.} Our evaluation focuses on applying heuristics to critique or improve existing dashboards. An acid test of \change{heuristics' utility} is their ability to productively shape the design process: we would ideally see how mindfulness of our heuristics impacts the final design of dashboards or the iterative process of \change{choosing} design alternatives. While we do note examples of participants saying that they would, as per $P25$, ``\textit{ apply this [heuristic] in my future dashboards},'' we leave this longitudinal assessment to future work. Additional future work is the connection of our heuristics to other forms of evaluation. For instance, does a \change{(re-) designed cooperative dashboard}
result in benefits to 
\change{user performance or satisfaction?}

\pheading{Extending the cooperative principles to other analytical workflows.} We believe that \textit{cooperative dashboards} represent a new perspective on visual analytics and potentially an emerging genre of visualization design. While it has been long understood that analytics is a \change{multi-stage process (e.g.}, the Pirolli/Card sensemaking loop~\cite{pirolli2005sensemaking}), there has been less work on 
visual analytics tools that operate \textit{across} stages. We consider dashboards
to be useful testbeds for learning about the structure of analytical conversations, and for testing novel designs to support users.
Cooperative dashboards allow a wide range of potential design or technique work for topics like mixed-initiative systems, NLIs, and rhetoric. \change{Beyond dashboards}, we also wish to apply these cooperative principles to other related forms (such as data stories) and media (such as designing visualizations for mobile or wearable devices).

\section{Conclusion}
In this paper, we explore the design of interactive dashboards as artifacts that support analytical conversations with their users. In particular, we explore how the role of language pragmatics and cooperative conversation can support data exploration, interaction, and reasoning. Inspired by existing models of conversational implicature and its states, we proposed and evaluated $39$ heuristics for helping guide the design of analytical conversation with interactive dashboards. These heuristics were iteratively validated with 16 visualization practitioners and subsequently evaluated by $52$ students to assess how useful they are for effectively authoring dashboards. Through the evaluation of these heuristics, we found that while dashboards tend to be effective at initiation and grounding of the conversation with the user, they struggle with other aspects of conversation that include turn-taking, repair \& refinement, and close. We hope that this work inspires the broader research and practitioner communities to explore new design and interaction paradigms for authoring more cooperative dashboard conversations.

\acknowledgments{
We thank the visualization researchers, practitioners, and students of Jio Institute, India, for their participation and feedback that helped inform the utility of this work. This research is also supported by 
NSF Award \#1900991 and The Roux Family Foundation.}
\bibliographystyle{abbrv-doi-hyperref}

\bibliography{main}

\end{document}